\documentclass[sigconf]{acmart}
\renewcommand\footnotetextcopyrightpermission[1]{} 
\AtBeginDocument{%
  \providecommand\BibTeX{{%
    Bib\TeX}}}

\usepackage{booktabs}  
\usepackage{subcaption} 

\usepackage{graphicx}
\usepackage{float}
\usepackage{stfloats}
\usepackage[ruled,linesnumbered]{algorithm2e}
\usepackage{algcompatible}
\usepackage{subcaption}
\usepackage{multirow}
\usepackage{makecell}
\usepackage{wrapfig}
\usepackage{xcolor,lipsum,subcaption}
\usepackage{enumitem}
\usepackage{amsmath}
\usepackage[frozencache,cachedir=minted-cache]{minted}

\def\BibTeX{{\rm B\kern-.05em{\sc i\kern-.025em b}\kern-.08em
    T\kern-.1667em\lower.7ex\hbox{E}\kern-.125emX}}
    
\begin{document}
\newcommand{\toolName}{EC-SpMV}
\newcommand{\formatName}{EC-CSR}

\title[{\toolName}]{Toward Efficient SpMV in Sparse LLMs via Block Extraction and Compressed Storage} 


\author{Junqing Lin}
\email{linjunqing@mail.ustc.edu.cn}
\orcid{0009-0007-1455-8725}
\affiliation{
    \institution{University of Science and Technology of China}
    \city{Hefei}
    \country{China}
}
\author{Jingwei Sun}
\authornote{Co-corresponding author}
\email{sunjw@ustc.edu.cn}
\orcid{0000-0001-5098-1503}
\affiliation{
    \institution{University of Science and Technology of China}
    \city{Hefei}
    \country{China}
}

\author{Mingge Lu}
\email{mingge@mail.ustc.edu.cn}
\affiliation{
    \institution{University of Science and Technology of China}
    \city{Hefei}
    \country{China}
}

\author{Guangzhong Sun}
\orcid{0000-0002-0794-7681}
\authornote{Co-corresponding author}
\email{gzsun@ustc.edu.cn}
\affiliation{
    \institution{University of Science and Technology of China}
    \city{Hefei}
    \country{China}
}



\begin{abstract}
Sparse Matrix-Vector Multiplication (SpMV) has become a critical performance bottleneck in the local deployment of sparse Large Language Models (LLMs), where inference predominantly operates on workloads during the decoder phase with a batch size of one. Existing SpMV kernels and sparse matrix formats, originally designed for scientific computing, fail to exploit the unique structure patterns inherent in sparse LLMs, resulting in suboptimal performance and excessive storage overhead.
This paper presents EC-SpMV, a GPU-optimized SpMV approach for accelerating sparse LLM inference. 
EC-SpMV introduces (1) a hierarchical block extraction algorithm that captures multiple granularities of block structures within sparse LLMs, and (2) a novel compressed sparse format (EC-CSR) that employs delta indexing to reduce storage overhead and enhance memory access efficiency.
Evaluated on real sparse weight matrices from LLaMA and OPT models, EC-SpMV achieves up to 6.44× speedup over state-of-the-art SpMV libraries and reduces storage overhead by up to 55.4\% compared to CSR. 

\end{abstract}

\maketitle

\section{Introduction}
Generative large language models (LLMs) have revolutionized natural language processing, achieving remarkable success in a variety of tasks \cite{zhao2023survey}.
While these models are typically deployed in data centers equipped with server-grade GPUs, there is a growing demand to deploy them locally on consumer-grade GPUs to address data privacy concerns and enable user-specific customization \cite{xia2023flash,song2023powerinfer,lyu2023llm,wang2023privatelora,yao2024survey}. 
However, local deployment on consumer-grade GPUs faces challenges due to limited storage capacity. For example, the Llama2-13b model \cite{touvron2023llama} requires 26 GB memory in FP16 precision, while consumer-grade GPUs, like NVIDIA GeForce RTX 4070, 4080, 4090, etc., typically offer only 8-24 GB memory. 

Neural network pruning \cite{xubesa, frantar2023sparsegpt,xia2023sheared,ma2023llm,sun2023simple,zhangdynamic,lu2024spp} has emerged as a promising approach to alleviate this challenge, effectively reducing both storage and computational overhead by removing redundant weights.
Pruning methods can be categorized by granularity into structured pruning \cite{xia2023sheared,ma2023llm} and unstructured pruning \cite{frantar2023sparsegpt,sun2023simple,zhangdynamic,lu2024spp}.
While unstructured pruning achieves higher sparsity without compromising accuracy, it results in sparse models that require specialized sparse computation kernels for efficient execution.

In the context of local inference, where a sparse LLM serves an individual user, Sparse Matrix-Vector Multiplication (SpMV) becomes the most critical kernel, appearing in essential operations such as QKV projections, output projections, and fully connected layers \cite{xia2023flash,song2023powerinfer}. 
The SpMV kernel computes the product of a sparse weight matrix $W$ and an input vector $X$ to produce an output vector $Y$. 
Despite its importance, optimizing SpMV performance in sparse LLMs faces two primary challenges: (1) the underutilization of data locality and (2) the inefficiency of sparse matrix storage formats.

\begin{figure}[h]
    \centering\includegraphics[width=0.45\textwidth]{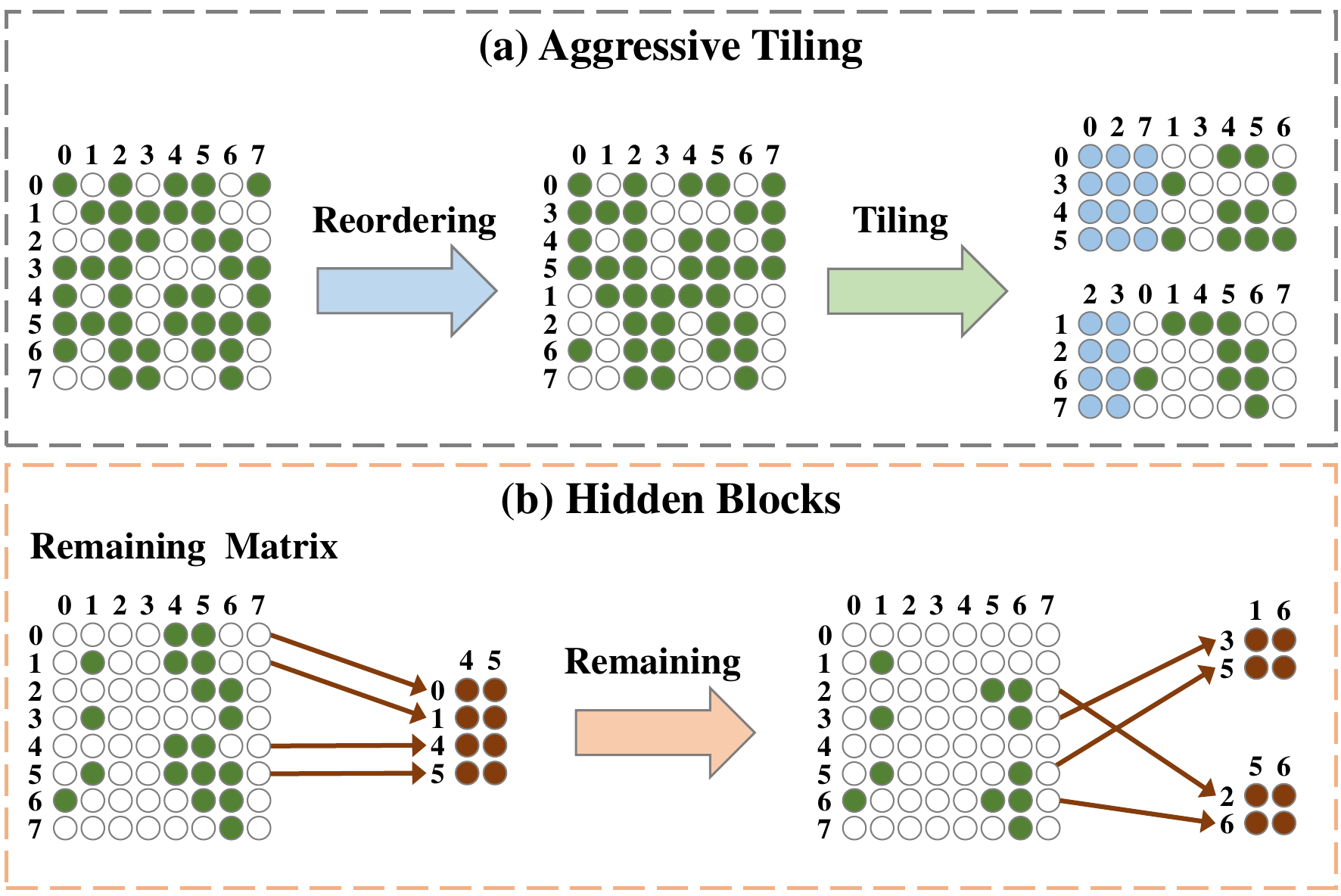}
    \caption{An illustration of aggressive tiling: blue and brown blocks indicate utilized and unutilized ones, respectively.}
    \label{fig:spmv}
\end{figure}

Data locality is fundamental for optimizing SpMV on GPUs.
Tiling techniques \cite{aktulga2014optimizing, niu2021tilespmv,  hong_adaptive_2019, jiang2020novel, du2022alphasparse, lin2024lo,zhao2025acc} have been proposed to partition the sparse matrix into smaller tiles to improve data reuse. 
Aggressive tiling methods \cite{hong_adaptive_2019, jiang2020novel, zhao2025acc} further improve data locality by reordering and clustering rows with similar non-zero distributions, enabling portions of the sparse matrix to be regarded as blocks and efficiently processed by dense computing, as illustrated in Figure 1(a).
These blocks also reduce the cost of indices since multiple elements within a block share one position index.
However, existing aggressive tiling methods fail to fully exploit potential blocks in sparse LLMs.
As illustrated in Figure \ref{fig:spmv}(b), hidden blocks within the sparse matrix can be revealed from two aspects: 
(1) extracting blocks from the sparse matrix and conducting further rounds of reordering and clustering on the remaining sparser matrix, 
and (2) identifying blocks with finer granularities, where granularity refers to the number of rows of a block.
To summarize, this suggests that achieving maximum data locality in sparse LLMs requires multiple rounds of block extraction at multiple levels of granularities, which is a combinatorial optimization problem.

Another critical challenge in optimizing SpMV lies in the sparse matrix storage format.
Existing formats \cite{xia2023flash,hong_adaptive_2019,jiang2020novel,gale_sparse_2020,guo2020accelerating,liu2015csr5,vazquez2011new,ashari2014fast,greathouse2014efficient}, originally designed for scientific computing, are ill-suited to the unique characteristics of sparse LLMs. These formats typically rely on absolute indices to represent the positions of non-zero elements, requiring high-precision data types (such as \texttt{uint16} or \texttt{uint32}) to handle the wide numerical range of column indices, leading to excessive storage and memory access costs.
Sparse weight matrices in sparse LLMs, however, exhibit moderate sparsity \cite{frantar2023sparsegpt,sun2023simple,zhangdynamic,lu2024spp} and a relatively uniform distribution of non-zero elements \cite{wilkinson2023register}, offering opportunities to optimize storage and memory access overhead.
Moreover, if we try to maximize data locality via multiple rounds of block extractions at multiple levels of granularities, existing formats do not match this design.
It motivates us to propose a novel format for the sparse matrices that contain highly hybrid structures.

To address these challenges, we propose {\toolName}, a novel approach that uses \textbf{E}xtraction and \textbf{C}ompression to accelerate \textbf{SpMV} kernels in sparse LLMs. 
First, {\toolName} introduces a hierarchical block extraction method that incorporates multi-round extractions of blocks at multi-level granularities to enhance data locality and also reduce index overhead. 
Since finding the optimal combination of blocks is an NP-hard problem \cite{kann1991maximum, karpinski2010computational}, our method is an approximate solution that greedily extracts and aggregates blocks from fine to coarse granularities in a reasonable time.
Second, {\toolName} introduces {\formatName} (\textbf{E}xtraction and \textbf{C}ompression-based \textbf{C}ompressed \textbf{S}parse \textbf{R}ow), a novel sparse matrix storage format designed specifically for sparse LLMs. 
{\formatName} separately maintains blocks of different granularities to ensure decoding efficiency, and leverages low-precision delta indexing to minimize storage and memory access costs. Besides, padding and permutation strategies are employed to optimize memory access efficiency.

We compare {\toolName} against five state-of-the-art SpMV implementations (cuSPARSE \cite{cusparse}, TileSpMV \cite{niu2021tilespmv}, CSR5 \cite{liu2015csr5}, AlphaSparse \cite{du2022alphasparse}, and DASP \cite{lu2023dasp}) on NVIDIA RTX 3080 Ti, RTX 3090, and RTX 4090, respectively. 
For FP32 precision, {\toolName} achieves up to $4.36 \times$ and a geometric average of $1.38 \times$ speedups over the best FP32 baseline AlphaSparse.
For FP16 precision, {\toolName} achieves up to $6.44 \times$ and a geometric average of $2.43 \times$ speedups over the best FP16 baseline DASP. Additionally, {\formatName} reduces storage overhead by 55.4\% compared to the CSR format.

Our key contributions are summarized as follows:
\begin{itemize}
    \item We propose a hierarchical block extraction method that optimizes data locality of SpMV in sparse LLMs.

    \item We propose the {\formatName} format for sparse weight matrices, which reduces storage overhead, enhances memory access performance, and preserves decoding efficiency.

    \item We conduct extensive experiments on three GPUs, demonstrating that our solution achieves better SpMV kernel performance and significantly reduces storage overhead compared to existing solutions.
\end{itemize}

\section{Background}

\subsection{LLM Inference}
LLMs primarily predominantly utilize the transformer architecture \cite{zhao2023survey}, which relies on self-attention mechanisms and fully connected layers.
The standard transformer performs a sequence of General Matrix Multiplication (GEMM) operations, including QKV projections, output projections, and feedforward layers.
Each GEMM operation multiplies a weight matrix $W$ of shape $(M, K)$ by an input matrix $X$ of shape $(K, N)$, producing an output matrix $Y$ of shape $(M, N)$. 
These GEMM operations account for the majority of latency during LLM inference \cite{xia2023flash}.

The inference process in LLMs consists of two phases: prefilling and decoding. During the prefilling phase, the model processes the input prompt in a single forward pass, computing hidden states for all input tokens simultaneously. In the autoregressive decoding phase, tokens are generated sequentially. 
The performance of LLMs is primarily constrained by the decoding stage \cite{xia2023flash,hong2024flashdecoding++}. In this phase, the model sequentially generates tokens by processing an input matrix $X$ of shape $(H, B)$, where $H$ is the hidden dimension and $B$ is the batch size.

\subsection{SpMV in LLM Inference}
In local deployments, LLMs typically process one request at a time, constraining the batch size $B$ to 1.
This results in the conversion of GEMM operations into General Matrix-Vector Multiplication (GEMV) operations.
After unstructured pruning, the weight matrices in LLMs become sparse, converting GEMV operations into SpMV operations. 
An SpMV operation multiplies a sparse weight matrix $W$ by an input vector $X$ to produce an output vector $Y$.

Sparse weight matrices in sparse LLMs exhibit moderate sparsity, as pruning can reduce model parameters by up to 70\% with minimal accuracy loss~\cite{frantar2023sparsegpt,sun2023simple,lu2024spp}. This level of sparsity is significantly lower than the over 99\% commonly observed in scientific computing applications. Moreover, the non-zero elements in sparse weight matrices tend to follow a uniform distribution~\cite{wilkinson2023register}.
These characteristics of sparse weight matrices can be exploited to optimize the performance of SpMV kernels in sparse LLMs.

\begin{figure}[b]
    \centering
    \includegraphics[width=0.48\textwidth]{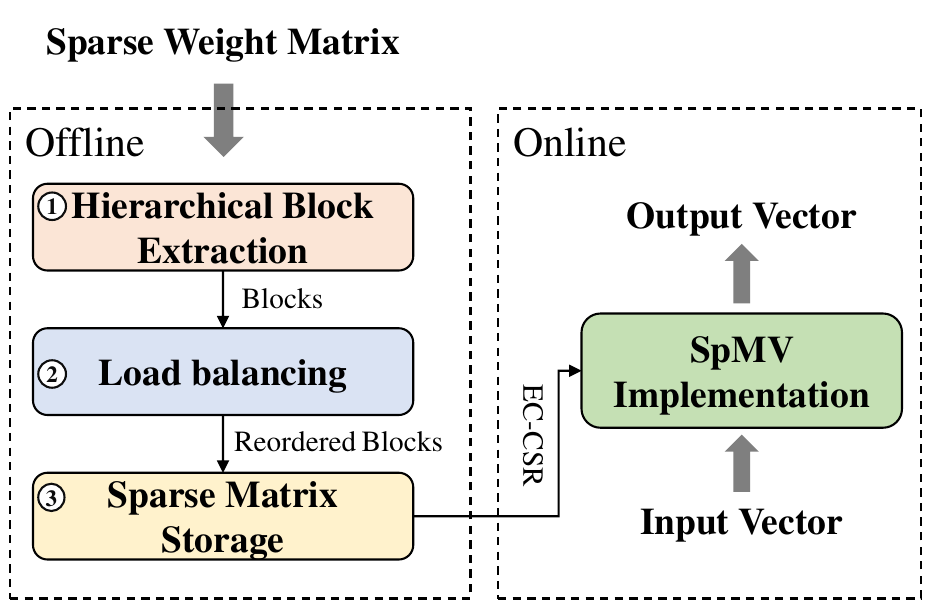}
    \caption{Overview of {\toolName}.}
    \label{fig:overview}
\end{figure}

\section{Overview of {\toolName}}

In this paper, we propose {\toolName} to improve the efficiency of SpMV in sparse LLMs, as illustrated in Figure~\ref{fig:overview}. Its two-phase design decouples the optimization process into an offline phase for sparse format generation, and an online phase for efficient computation.

In the offline phase, {\toolName} adopts a novel hierarchical block extraction method that systematically identifies blocks across multiple granularities, ranging from fine to coarse  (Section \ref{sec:hbe}). 
At each granularity level, {\toolName} performs multiple rounds of extraction to maximize block coverage.
The hierarchical block extraction improves memory access efficiency by improving data locality and reducing redundant index accesses.
The extracted blocks are subsequently clipped and reordered to balance the workload for the SpMV kernel (Section \ref{sec:lb} ). Finally, they are stored in a specialized sparse format, {\formatName}, which reduces storage memory access overhead compared to existing formats (Section \ref{sec:sms}).  
The use of {\formatName} also ensures efficient memory coalescing and vectorized memory access, making the SpMV operation highly optimized for GPU execution.

In the online phase, the SpMV kernel, implemented based on the {\formatName} format, performs matrix-vector multiplication by combining the input vector with the stored blocks to produce the output vector (Section \ref{sec:impl}). 

The following sections provide detailed explanations of each phase.

\section{Hierarchical Block Extraction}
\label{sec:hbe}
Sparse matrices often exhibit local block structures that can be exploited to improve memory efficiency and data reuse in SpMV. Grouping non-zero elements into blocks reduces indexing overhead and input vector accesses.
We define a \textbf{\(g\)-grained block} as a set of elements within a submatrix comprising \(g\) rows and \(n\) columns (not necessarily contiguous). 
Elements within the same block column share a column index,  requiring only a single input vector access per block column.
As a result, the amortized access cost per element is reduced to \(\frac{L}{g}\), where \(L\) denotes the cost of accessing one input vector element. The total memory access cost for input vector in an SpMV kernel is then given by:
\[
 \sum_{e \in \text{non-zero}(A)} \frac{L}{g(e)} ,
\]
where \(A\) is the sparse matrix and \( g(e) \) denotes the granularity of the block containing element \( e \).
The objective of block extraction is to maximize the granularity \( g(e) \) across all elements, thereby minimizing the overall memory access cost.

However, current aggressive tiling methods~\cite{hong_adaptive_2019, jiang2020novel, zhao2025acc}, which cluster rows into row groups, exhibit significant limitations in block extraction from sparse matrices. 
Specifically, these methods impose a fixed granularity on each row group, restricting the identification of fine-grained blocks within each group.
Moreover, they fail to exploit the opportunity to reorganize the remaining sparse segments into new row groups.

To overcome these limitations, we propose a hierarchical block extraction method that supports adaptive and multi-granularity block identification. {\toolName} performs multi-level aggregation, where each level generates coarser-grained blocks from finer-grained ones extracted at the previous level. 
This hierarchical strategy ensures that fine-grained elements are progressively integrated into as large a block as possible.
Additionally, {\toolName} employs multi-round extraction, allowing different non-zero elements within a row to participate in multiple blocks, thereby maximizing block coverage at each level.

\begin{figure*}[h]
    \centering
    \includegraphics[width=1.0\textwidth]{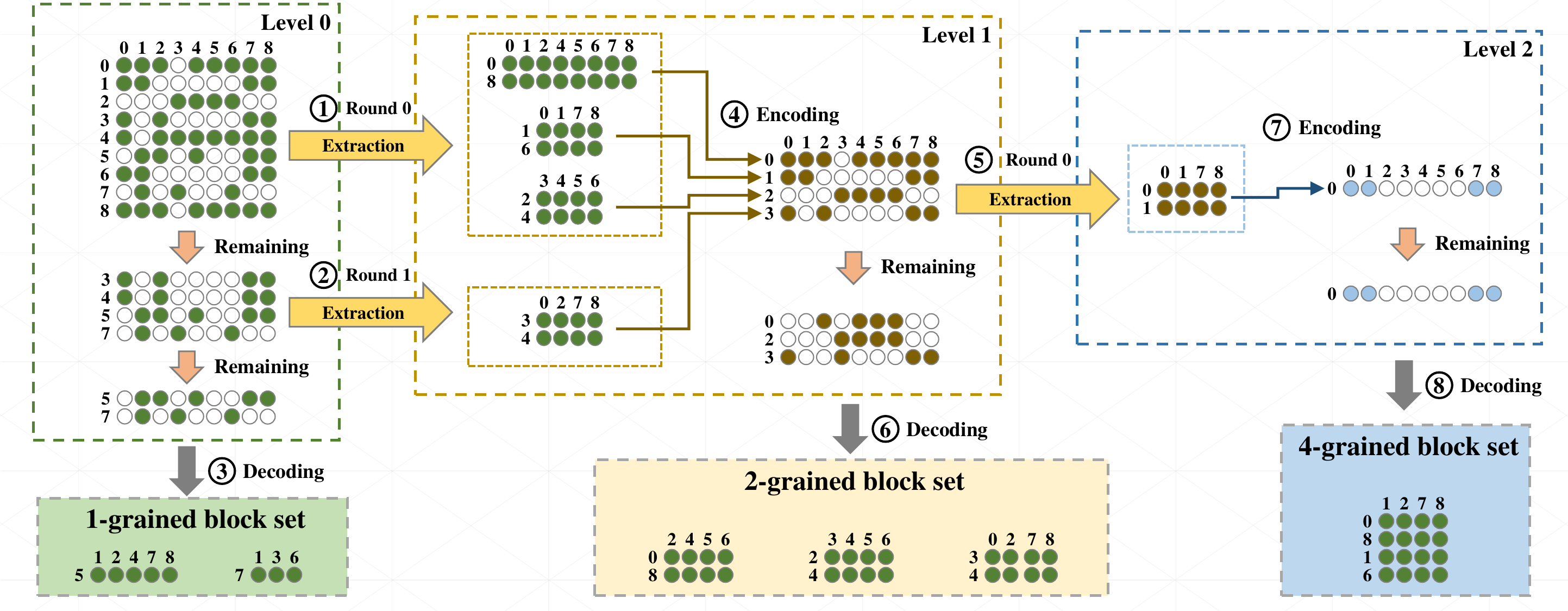}
    \caption{ An illustration of hierarchical block extraction. 
    Each non-zero element in a sparse matrix represents a dense column of size $2^{Level} \times 1$. In other words, a green, brown, blue element indicates a $1 \times 1$, $2 \times 1$, $4 \times 1$ dense column, respectively.
    }
    \label{fig:extracting}
\end{figure*}

\begin{algorithm}[h]
    \small
    \caption{Hierarchical Block Extraction}
    \label{alg:hbe}
        \SetKwInOut{Input}{Input}\SetKwInOut{Output}{Output}
        \Input{ 
                Sparse Matrix, \quad \textbf{$A$} \\
        }
        \Output{ 
                Block sets at various granularities, \quad  \textbf{$block\_sets$} \\
        }    
        $block\_sets = []$\;
        $A_E = A$\;
        
        \tcc{multi-level aggregation}
        
        \While{$True$}
        {
        $extracted\_blocks =  []$\; 
        $A_R = A_E$\;
        \tcc{multi-round extraction}
        \While{$True$}
        {
            
            $row\_pairs = rowMatching(A_R)$\;
            $blocks_{R}, A_R = extractBlock(A_R, row\_pairs)$\;
            \If{$blocks_{R}.empty()$}{
                $break$\;
            }
            $extracted\_blocks.append(blocks_{R})$\;
            
        }
        $A_{E} = encode (extracted\_blocks)$\;

        $decoded\_blocks = decode(A, A_{R})$\;        
        $block\_sets.append(decoded\_blocks)$\;   
        
        \If{$extracted\_blocks.empty()$}{
            $return\ block\_sets$\;
        }

        }

\end{algorithm}

\subsection{Algorithm Overview}

Algorithm \ref{alg:hbe} outlines the process of hierarchical block extraction, while Figure \ref{fig:extracting} illustrates an example of the extraction.
At each level, repeated extraction rounds yield as many blocks as possible (lines 6-13).
During each round, {\toolName} identifies the most similar row to each row to form pairs (line 7) and extracts common columns within pairs to construct 2-grained blocks (line 8). 
The extracted elements are set to zero, and the remaining elements are extracted in the next round.
The multi-round extraction process halts when no further usable blocks can be extracted (lines 9-11).
These blocks are then encoded into a new sparse matrix (line 14), where each element is an aggregation of two elements from the sparse matrix of the previous level. 
This matrix $A_E$ facilitates further extractions at progressively coarser granularities in subsequent levels.
Meanwhile, the remaining sparse matrix $A_{R}$ is then converted into blocks, which is the output of current granularity (lines 15-16).

Next, we introduce detailed motivations and designs of multi-level aggregation and multi-round extraction, respectively.

\subsection{Multi-level Aggregation}
Sparse matrices in sparse LLMs inherently exhibit block patterns at various granularities. 
Coarse-grained blocks, which span more rows, are particularly advantageous as they enhance data reuse and reduce indexing overhead. 
However, focusing solely on coarse-grained risks neglects fine-grained structures that also remain valuable for optimizing data reuse.
To fully exploit the reuse potential across all granularities, a holistic block extraction strategy is necessary.

A straightforward approach might extract blocks sequentially, starting from coarse-grained blocks and progressing to finer granularities.
However, for a sparse matrix with $M$ rows and a target block granularity $G$, there are $\binom{M}{G}$ possible row groupings.
Given that $G$ is often much smaller than $M$, the combination space grows exponentially at coarser granularities, resulting in computationally expensive extraction, especially when multiple granularities are considered simultaneously. 
The total number of possible extraction combinations across granularities is given by:
\begin{equation}
 \sum_{G \in Gs} \binom{M}{G} \times R^G,    
\end{equation}
where $Gs$ denotes the set of considered granularities and $R^G$ is the extraction count at granularity \(G\).
Moreover, a coarse-grained block comprises multiple finer-grained units. 
Moreover, a coarse-grained block are composed of multiple finer-grained units. This hierarchical structure introduces a dilemma: extracting fine-grained blocks first may eliminate the opportunity to extract larger blocks later, while extracting coarse-grained blocks first requires reconsider the composition of fine-grained blocks, leading to redundancy and inefficiency.

To address this, {\toolName} adopts a multi-level aggregation approach that incrementally constructs coarser blocks from finer-grained blocks.
This bottom-up strategy avoids redundant extraction and enables efficient reuse of intermediate results across granularities.
As described in Algorithm \ref{alg:hbe}, {\toolName} performs multi-round extraction to identify all 2-grained blocks from the current sparse matrix $A_R$ at each level (lines 5–13). 
Each extracted block is then encoded as a sparse vector of length $K$, which corresponds to the number of columns in $A_R$ (line 14). 
Each non-zero element in this vector represents a specific column within the block, where the column length defines the block's granularity.
The collection of these vectors constitutes a new sparse matrix of dimensions $(M^G, K)$, where $M^G$ is the number of extracted blocks. 
For example, at level 1 in Figure \ref{fig:extracting}, four 2-grained blocks are encoded into a new sparse matrix of shape (4, 9). 
This newly constructed matrix serves as the input for the next aggregation level, enabling the extraction of coarser-grained blocks whose granularity is double that of the current level.
Meanwhile, the rows in the remaining sparse matrix are decoded into blocks and added to the current-level block set (lines 15–16). 
For example, at level 1 in Figure~\ref{fig:extracting}, three non-empty rows remain in \(A_R\) after multi-round block extractions. Each non-empty row is decoded into a 2-grained block, with the corresponding block column indices aligning with indices of the non-zero elements in the row.

Compared to an exhaustive enumeration, whose cost is represented as equation (1), this approach significantly reduces computation by limiting the number of combinations across granularities:
\begin{equation}
 \sum_{G \in Gs} \binom{M^{G}}{2} \times R^G,   
\end{equation}
where $M^{G}$ denotes the number of blocks extracted from the preceding granularity level.

\subsection{Multi-round Extraction}

In sparse matrices, different non-zero elements within a single row may participate in multiple distinct block structures after properly reordering.
For example, in the level 0 in Figure \ref{fig:extracting}, row 4 shares non-zero positions with row 2 at columns 3, 4, 5, and 6, and with row 3 at columns 0, 2, 7, and 8.
In this example, row 4 is divided into two parts, each forming a block with row 2 and row 3, respectively.
Existing aggressive tiling methods restrict each row to contribute to a single block, thereby limiting block coverage.

To address this limitation, {\toolName} adopts a multi-round extraction strategy that allows non-zero elements within a row to be distributed across multiple blocks over successive rounds.
In each extraction round, {\toolName} first computes row-wise similarity, defined by the ratio of shared columns between two pairs, and pairs each row with the one having the highest similarity. 
It then identifies and extracts the shared columns between paired rows, forming a two-row block (i.e., a 2-grained block).  
Once a round completes, the non-zero elements included in extracted blocks are marked as zero, ensuring they are excluded from subsequent rounds. This process iterates over the remaining uncovered elements until no additional blocks can be identified.
By zeroing out previously extracted elements after each iteration, the algorithm incrementally reduces the sparse matrix into smaller subproblems. 
This facilitates the identification of additional useful blocks in subsequent rounds. Figure~\ref{fig:extracting} illustrates the extraction process at level 0: three blocks are initially extracted, followed by a second round applied to the residual matrix, which yields an additional block spanning four columns.

\begin{algorithm}[]
    \small
    \caption{Row Matching}
    \label{alg:rowmatching}
    \SetKwInOut{Input}{Input}\SetKwInOut{Output}{Output}
    \Input{ Sparse Matrix, \quad \textbf{$A$}\\
    }
    \Output{ 
            Matching Row Pairs ,\quad \textbf{$pairs$} \\
    }       
    $W = getSimilarityMatrix(A)$\;
    $unselected\_rows = A.rows$\;
    $pairs = []$\;
    \While{$unselected\_rows.empty() == False$}
    {
        $row = unselected\_rows.pop()$\;
        $matched\_row = argmax(W[row][unselected\_rows])$\;
        $pairs.append((row, matched\_row))$\;
        $unselected\_rows.remove(matched\_row)$\;
    }
    
    $return\ \ \ \ pairs$\;
    
\end{algorithm}

The primary challenge in each extraction round lies in effectively clustering similar rows to maximize block coverage. Specifically, our hierarchical method aims to extract 2-grained blocks at each granularity level, where each block comprises two rows from the previous extraction level. 
Therefore, the objective is reduced to identifying an optimal matching row for each row.
This problem can be formulated as a maximum weight matching on a complete graph, where nodes represent rows and edge weights correspond to the number of shared columns between row pairs. 
The goal is to identify a matching that maximizes the total weight across all matched pairs.

The exact algorithm for maximum weight matching in complete graphs, as proposed by \cite{duan2018scaling}, has a time complexity of $O(n^{2.5} \log(nN))$, where $n$ denotes the number of nodes and $N$ is the maximum edge weight. 
This becomes computationally impractical for large-scale scenarios, such as those involving sparse matrices with tens of thousands of rows.
Meanwhile, in sparse LLMs with uniformly distributed sparsity and many similar row pairs, precise matching is unnecessary.
To balance efficiency and effectiveness,  we adopt a greedy algorithm, as detailed in Algorithm \ref{alg:rowmatching}. Initially, the algorithm counts the number of overlapping columns between rows (line 1). For each unselected row, it iterates through the unmatched rows (line 5) and selects the unmatched row with the highest overlap as the optimal match (lines 6–7). The matched row is then marked as selected to avoid duplicate assignments (line 8). This process continues until all rows are matched or no further pairs can be formed.
While the greedy algorithm does not guarantee optimality, it significantly reduces computational complexity to $O(n^2 \log n)$, making it feasible for large-scale sparse matrices.

\section{Load Balancing}
\label{sec:lb}
In addition to hierarchical block extraction, achieving efficient SpMV performance also requires balanced workload distribution across GPU cores. 
To optimize hardware resource utilization, {\toolName} tiles the SpMV kernel workload into multiple tiles. On GPUs, each thread block processes multiple blocks, with each warp handling a single block. 
The workload within each warp is evenly distributed across the threads to ensure load balancing. 
However, variations in block sizes can disrupt load balancing both within each thread block and across streaming multiprocessors (SMs) in a GPU.
Load imbalance may result in the underutilization of computational resources on a GPU, causing performance degradation.

To address this issue, {\toolName} employs two techniques: block clipping and reordering. 
Block clipping divides long blocks into shorter segments using a dynamically adjusted threshold.
After block clipping, the resulting short blocks are independently assigned to warps, ensuring that no warp is overloaded with excessively large computation.
Our reordering strategy, modified from the row-swizzle method \cite{gale_sparse_2020}, enhances load balancing in the SpMV kernel. Specifically, we first sort blocks within each block set in descending order of nnz, then sort the block sets themselves by decreasing granularity.

These block clipping and reordering methods promote a more uniform distribution within each thread block. Additionally, GPU scheduling typically dispatches the initial wave of blocks in a round-robin fashion across SMs, followed by dispatching the remaining blocks based on SM execution time \cite{gale_sparse_2020}. Therefore, these methods also enhance load balancing across the SMs.

\section{Sparse Matrix Storage}
\label{sec:sms}
The storage format of sparse matrices is critical for both reducing memory footprint and achieving high-performance inference in sparse LLMs.
Existing formats \cite{niu2021tilespmv,liu2015csr5,vazquez2011new,ashari2014fast,greathouse2014efficient,bell2008efficient,merrill2016merge}, while well-suited for scientific computing scenarios, exhibit limitations in the context of sparse LLMs.
These limitations primarily arise from two factors: (1) the excessive storage overhead incurred by absolute indexing, 
and (2) the inefficiency in handling irregular blocks within sparse LLM weight matrices, where blocks vary in size and consist of scattered, non-contiguous elements in the original matrix layout.
To overcome these limitations, we design {\formatName}, a sparse matrix storage format tailored to the block structures extracted in our pipeline.
{\formatName} is optimized for hierarchical block sets, enabling efficient index compression, memory coalescing, and vectorized access on modern GPUs.

\begin{figure}[h]
    
    \includegraphics[width=0.48\textwidth]{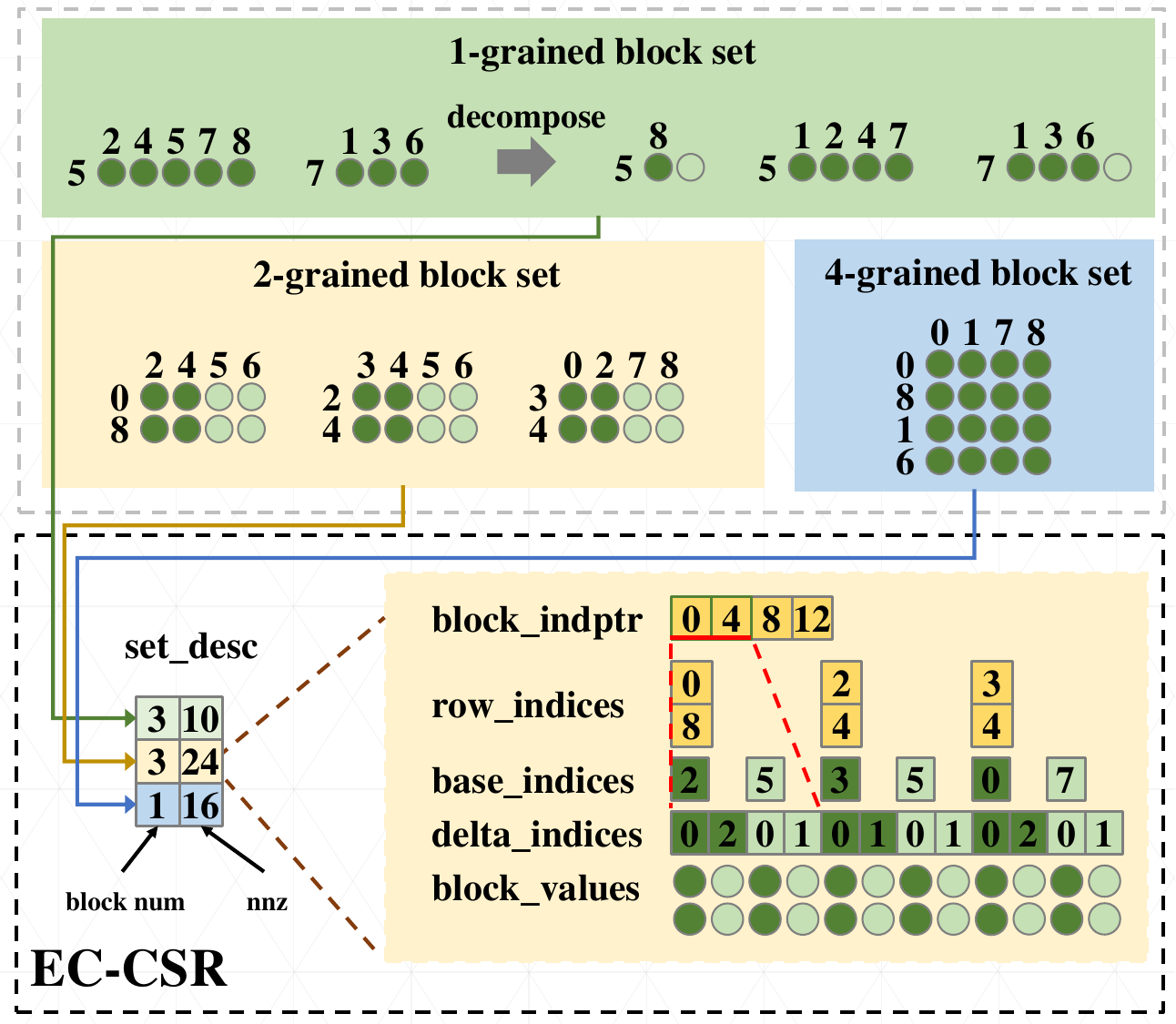}
    \caption{An example of {\formatName} storing three block sets. Assume that the vector type used is of size 2 and the warp\_size is 2. The detailed storage structure of the 2-grained block set is zoomed in. The dark green cells in $base\_indices$, $delta\_indices$, and $block\_values$ are assigned to thread 0, and the light green cells are assigned to thread 1.}
    \label{fig:format}
\end{figure}

\subsection{Hierarchical Block Representation}
The hierarchical block extraction process (Section \ref{sec:hbe}) generates block sets at multiple granularities, which are then reordered using load balancing techniques (Section \ref{sec:lb}). 
As shown in Figure~\ref{fig:format},  {\formatName} represents each reordered block set using five arrays:  \textbf{row\_indices}, \textbf{block\_indptr}, \textbf{base\_indices}, \textbf{delta\_indices}, and \textbf{block\_values}. 
Each block set is also associated with metadata \textbf{set\_desc}, describing the number of blocks and the total number of non-zero elements.
Additionally, the 1-grained block set requires decomposition to enable efficient vectorization, as detailed in Section \ref{sec:mao}.

The following outlines the process of transforming reordered blocks into the {\formatName} format.
Each block is processed sequentially: 
The block is first divided into sub-blocks of size \texttt{warp\_size}, and index compression is applied to obtain the base and delta indices. 
Meanwhile, these sub-blocks are padded to satisfy memory address alignment.
The base indices are appended to \textbf{base\_indices}, while the delta indices and block values are permuted to optimize memory access patterns. These permuted arrays are stored in \textbf{delta\_indices} and \textbf{block\_values}, respectively. 
Finally, the block offset is appended to \textbf{block\_indptr}, and the corresponding row indices are added to \textbf{row\_indices}.
Further details on index compression and memory access optimization are discussed in subsequent sections.

\subsection{Index Compression}
Following block extraction, a critical step in minimizing storage overhead is the encoding of non-zero element positions. Traditional sparse matrix formats~\cite{xia2023flash,hong_adaptive_2019,jiang2020novel,gale_sparse_2020,guo2020accelerating} typically use absolute indexing, where each non-zero stores its full column index. However, in the context of LLMs, sparse weight matrices typically exhibit moderate sparsity~\cite{frantar2023sparsegpt,sun2023simple} and a relatively uniform distribution of non-zero elements~\cite{wilkinson2023register}.
These characteristics often lead to locally clustered indices, making absolute indexing both redundant and inefficient.

To address this limitation, we propose a delta indexing scheme that exploits the spatial locality of non-zero elements. 
Instead of storing absolute positions, we encode the numerical differences between consecutive indices within each block. 
These delta values typically fall within a narrow range, enabling compact, low-precision encoding and significantly reducing storage overhead. 
Moreover, delta indexing minimizes memory access overhead during SpMV computations by leveraging these compact delta values, thereby reducing the number of memory transactions required.

\begin{figure}[h]
    \centering \includegraphics[width=0.4\textwidth]{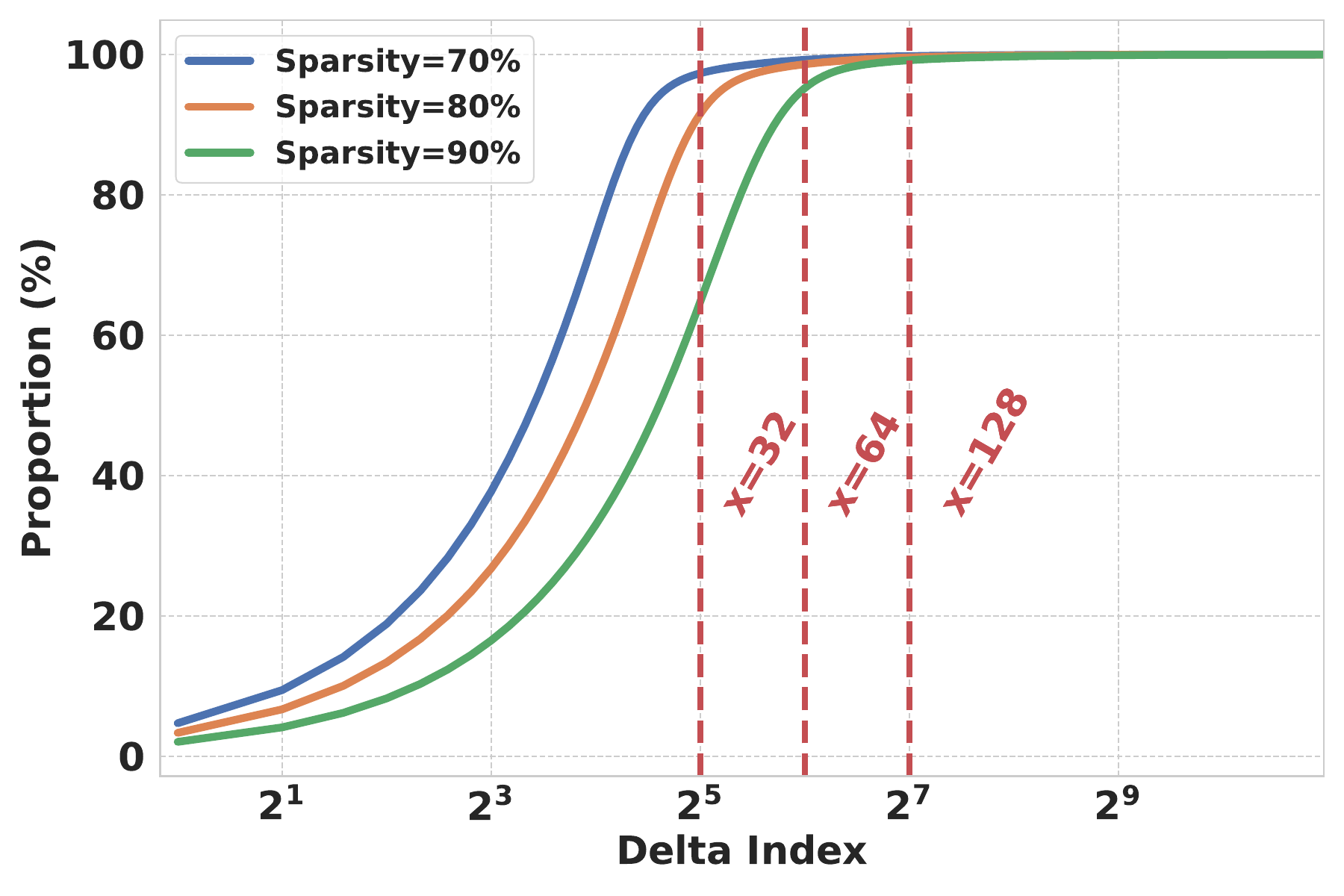}
    \caption{Cumulative distribution of delta indices in sparse weight matrices.}
    \label{fig:delta_index}
\end{figure}

Figure \ref{fig:delta_index} illustrates the cumulative distribution of delta indices in the sparse weight matrices of Llama2 models at scales of 7B, 13B, and 70B, with sparsity levels of 70\%, 80\%, and 90\%, respectively.
The results indicate that most delta indices remain below specific thresholds: 32 at 70\% sparsity, 64 at 80\% sparsity, and 128 at 90\% sparsity. 
Consequently, low-precision storage for these delta indices is feasible, significantly reducing the storage overhead compared to using absolute indices. 

However, Figure \ref{fig:delta_index} also indicates that a small proportion of delta indices fall outside the threshold, referred to as outliers. Storing all delta indices with higher precision to account for these outliers introduces unnecessary storage overhead. 
To address this, we utilize a predefined precision range \( R_P \) and extract only adjacent columns within this range during block extraction. For 1-grained blocks, zero elements are inserted at intervals of $R_P - 1$ between widely spaced non-zero elements, ensuring that delta indices remain within the desired range.

During execution, each block is processed by a warp, which contains \texttt{warp\_size} threads. 
Thus, we partition a block into \texttt{warp\_size} sub-blocks. 
The $i$-th subblock handles the columns within the range $\left[i \cdot \frac{\text{nnc}}{\text{warp\_size}}, (i+1) \cdot \frac{\text{nnc}}{\text{warp\_size}}\right)$, where \texttt{nnc} denotes the number of non-empty columns in the block.
To improve the efficiency of index decoding, each thread maintains an base index and the corresponding $\frac{\text{nnc}}{\text{warp\_size}}$ delta indices of the columns it processes. 
For example, in the first block of the 2-grained block set illustrated in Figure \ref{fig:format}, thread 0 (dark green) holds a base index of 2 and delta indices of 0 and 2, and thread 1 (light green) holds an base index of 5 and delta indices of 0 and 1. 
The collection of all base indices in the block set forms the \texttt{base\_indices} array, while all delta indices form the \texttt{delta\_indices} array.
Each thread in the index decoding process first retrieves the base index \( I_0 \) and sequentially processes each column. The index of the column \( I_k \) is computed as:
\[
I_k = I_0 + \sum_{i=1}^{k} \Delta I_i
\]
where \( \Delta I_i \) represents the delta index of the $i$-th column.

\subsection{Memory Access Optimization}
\label{sec:mao}
Inefficient memory access can cause high latency and suboptimal throughput, thereby limiting the performance of SpMV kernels. 
To address this, {\formatName} explicitly supports vectorized and coalesced memory access patterns, which are essential for maximizing memory throughput and minimizing access overhead on modern GPUs. 
These optimizations enhance SpMV performance by reducing memory instructions and improving bandwidth utilization. This section details the implementation of these memory access optimizations within {\formatName}.

\subsubsection{Vectorized memory access}
Vectorized memory access enables the simultaneous processing of multiple data elements per thread, reducing the number of memory instructions. For example, using the \texttt{float4} datatype enables a single memory operation to read or write four floating-point numbers at once, allowing each thread to execute one memory instruction instead of four.
However, the effectiveness of vectorized memory access depends heavily on data alignment and layout, which can be challenging in sparse formats. To address this, we employ two strategies to ensure alignment.
First, we extract only multiples of \(warp\_size \times vector\_size\) columns during block extraction, where \(warp\_size\) and \(vector\_size\) denote the number of threads per warp and the size of vector type, respectively. 
This ensures that each thread for extracted blocks processes non-empty columns in multiples of \(vector\_size\) to leverage vectorized operations.
Second, each 1-grained block is partitioned into a \textit{long block}—comprising a multiple of \(warp\_size \times vector\_size\) non-zero elements to enable efficient vectorized operations—and a \textit{short block} containing the remaining elements. To ensure alignment, short blocks are padded to the nearest multiple of \(warp\_size\). 
As illustrated in Figure~\ref{fig:format}, the first 1-grained block is split into two sub-blocks of sizes 1 and 4, with the short block padded to size 2.
These strategies facilitate vectorized access for most non-zero elements with minimal padding overhead.

\begin{figure}[h]
    \centering \includegraphics[width=0.45\textwidth]{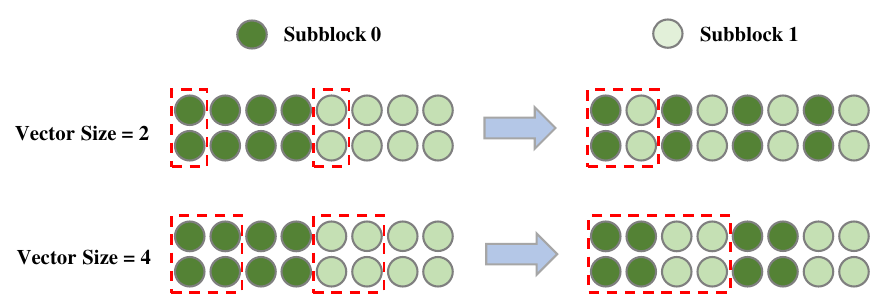}
    \caption{Illustration of permutation. Assume \texttt{warp\_size} is 2.}
    \label{fig:permute}
\end{figure}

\subsubsection{Coalesced memory access}
Coalesced memory access occurs when all threads in a warp simultaneously access consecutive memory addresses, minimizing the number of memory transactions. 
To achieve this, each block is partitioned into \texttt{warp\_size} sub-blocks, with each subblock assigned to a corresponding thread within the warp.
If each subblock is stored contiguously, as shown on the left of Figure~\ref{fig:permute}, threads accessing delta indices or non-zero values would retrieve data from non-contiguous memory addresses, leading to inefficient memory transactions.
To address this issue, we permute the $delta\_indices$ and $block\_values$ arrays. 
Specifically, the delta indices and non-zero values for each thread are divided into segments of size \texttt{vector\_size}, representing the number of elements each thread reads at a time. These aligned segments are then merged across threads into chunks of size $\texttt{warp\_size} \times \texttt{vector\_size}$ and arranged sequentially. 
Figure~\ref{fig:permute} illustrates examples with vector sizes of 2 and 4, where elements are arranged in column-major order.

\begin{listing}[h]
\begin{minted}[frame=single, framesep=2mm, fontsize=\footnotesize, escapeinside=||, linenos, keywordcase=upper]{cuda}
/* Input: 
    Block metadata array: block_sets
    Input vector values: x_values
   Output: 
    Output vector values: y_values
*/

// Compute global warp ID
warp_id = threadIdx.x / warp_size;
lane_id = threadIdx.x % warp_size;
warp_id_global = blockIdx.x * (gridDim.x/warp_size) + warp_id;

// Extract block-related data
G, nnc, row_indices, base_indices, delta_indices, block_values = 
    getBlockData(block_sets, warp_id_global);

// Define thread-level parameters
load_size = warp_size * vector_size;
num_iterations = nnc / load_size;
delta_indices += num_iterations * vector_size;
block_values += num_iterations * vector_size * G;
base_index = base_indices[0];


res[0:G] = 0;

// Prefetch first batch of values
next_deltas = delta_indices[0];
next_values = block_values[0];

// Iterate over non-empty columns in the subblock
for (int i = 0; i < num_iterations; i++) {
    current_deltas = next_deltas;
    current_values = next_values;

    if (i < num_iterations - 1) {
        next_deltas = delta_indices[load_size * (i + 1)];
        next_values = block_values[load_size * (i + 1)];
    }

    // Perform SpMV computation
    for (int j = 0; j < vector_size; j++) {
        base_index += current_deltas[j];
        x_value = x_values[current_index];

        // Accumulate partial results for each row in the subblock
        for (int k = 0; k < G; k++) {
            res[k] += current_values[j * G + k] * x_value;
        }
    }
}

// Perform warp-level reduction to aggregate results
for (int j = 0; j < g; j++) {
    results[j] = warpReduceSum(res[j]);
}

// Store final results using atomic addition
if (lane_id == 0) {
    for (int j = 0; j < g; j++) {
        atomicAdd(y_values + row_indices[j], res[j]);
    }
}
\end{minted}
\caption{SpMV Implementation using {\formatName}
}
\label{alg:kernel_impl}
\end{listing}

\section{SpMV Implementation}
\label{sec:impl}

We implement the SpMV GPU kernel using the EC-CSR format, as detailed in Listing \ref{alg:kernel_impl}. The implementation is designed to maximize parallelism and minimize memory access overhead by leveraging warp-level operations and double buffering.

The algorithm begins by calculating \( warp\_id\_global \), which identifies the current warp (lines 9-11).
Each warp is assigned a distinct block to process in parallel.
The block attributes, including the number of rows $G$, the number of non-zero columns $nnc$, the row indices, base column indices, delta indices, and block values, are retrieved for the assigned block (line 14).

To efficiently load data, \( \texttt{warp\_size} \times \texttt{vector\_size} \) elements are processed by each warp in each iteration, where \texttt{vector\_size} is set to 4 for most blocks but is reduced to 1 for 1-grained blocks.
The number of iterations per thread is determined by the block, ensuring that the warp handles a balanced workload (lines 18-19).
Before entering the main loop, each thread loads its base column index, labeled \( current\_index \) (line 22). 

\begin{figure*}[b]
    \centering
    \begin{subfigure}[]{0.33\textwidth}
        \centering
        \includegraphics[width=\textwidth]{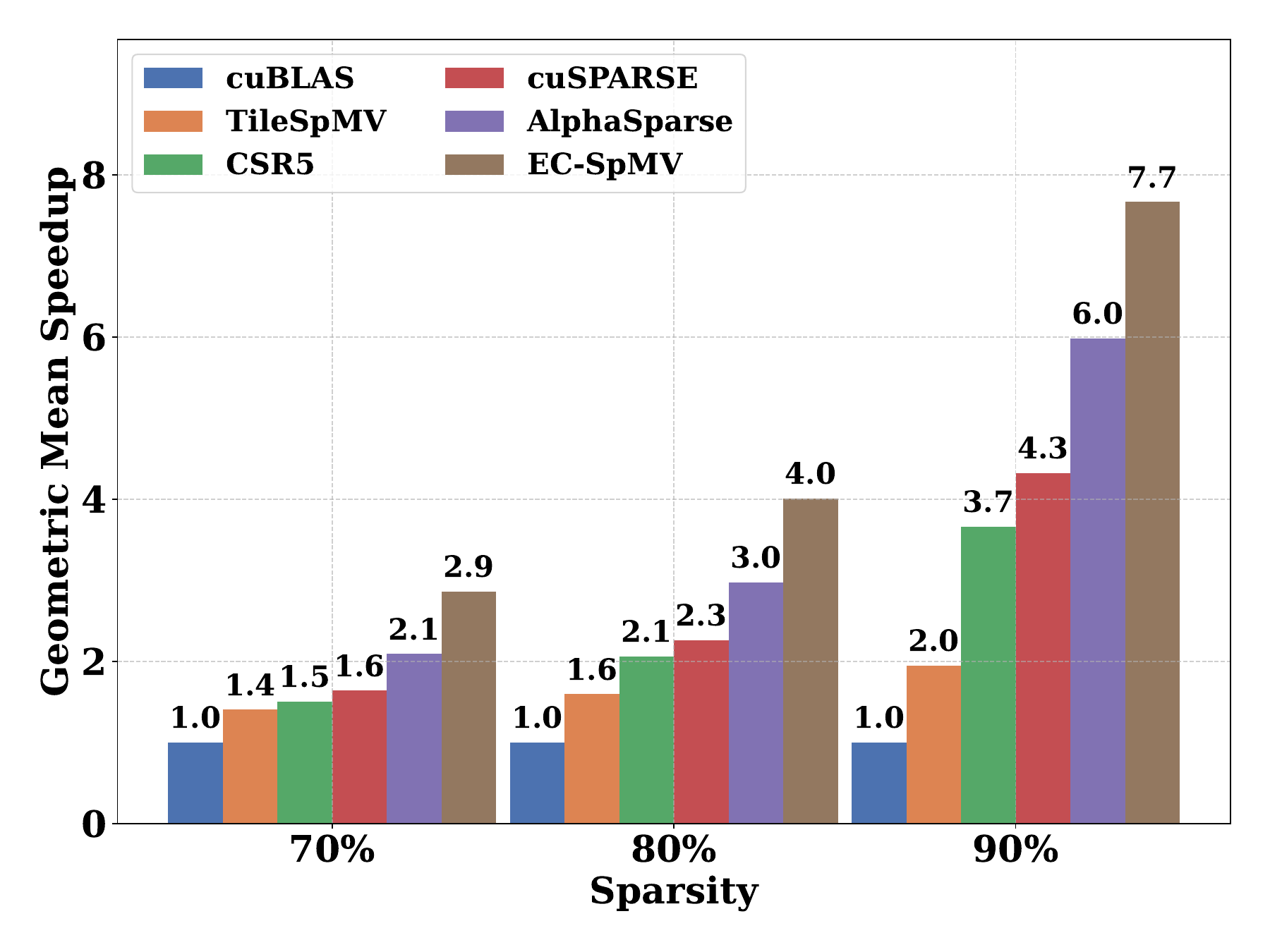}
        \caption{3080 Ti for FP32 precision}
    \end{subfigure}
    \begin{subfigure}[]{0.33\textwidth}
        \centering
        \includegraphics[width=\textwidth]{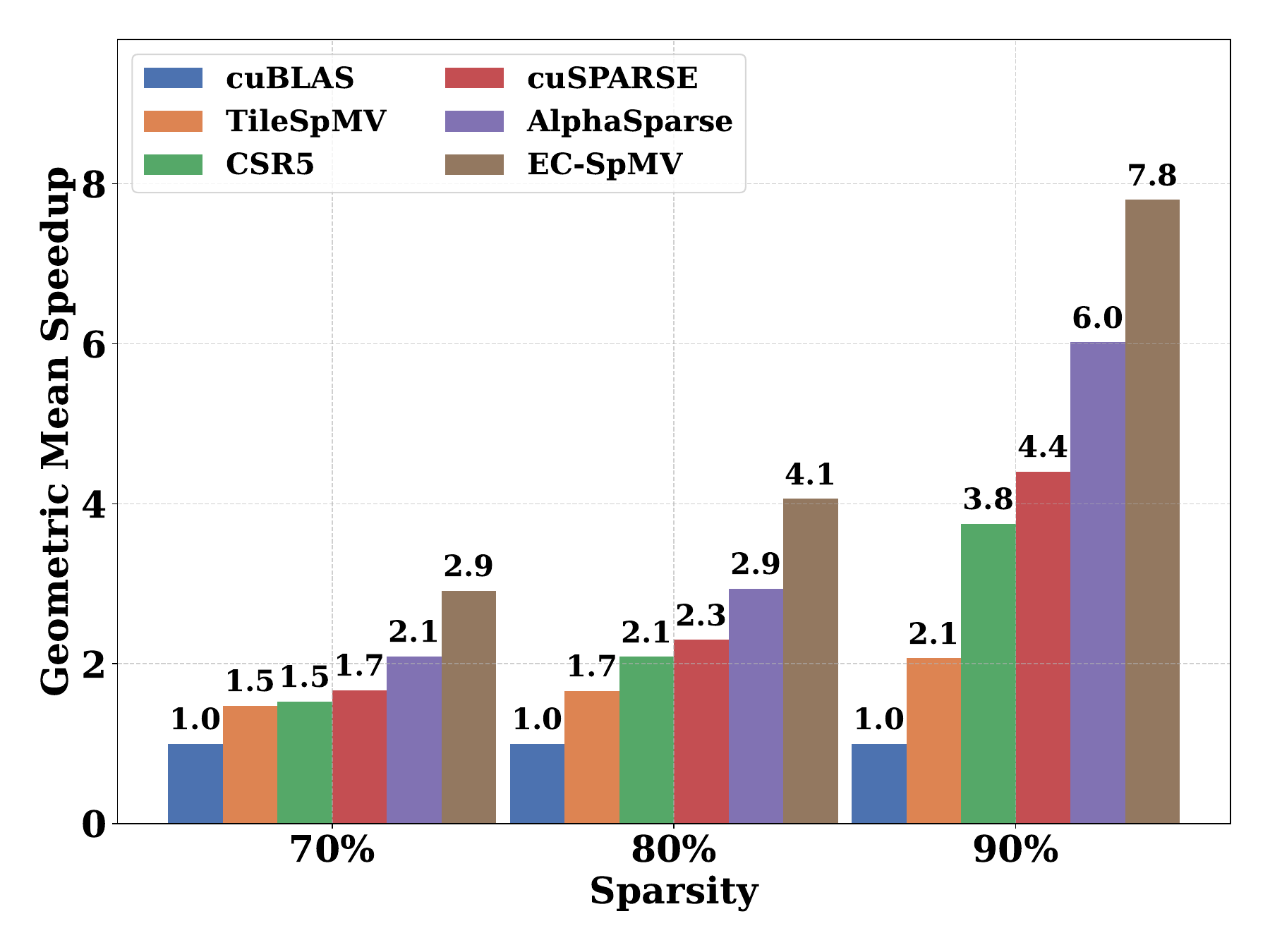}
        \caption{3090 for FP32 precision}
    \end{subfigure}
    \begin{subfigure}[]{0.33\textwidth}
        \centering
        \includegraphics[width=\textwidth]{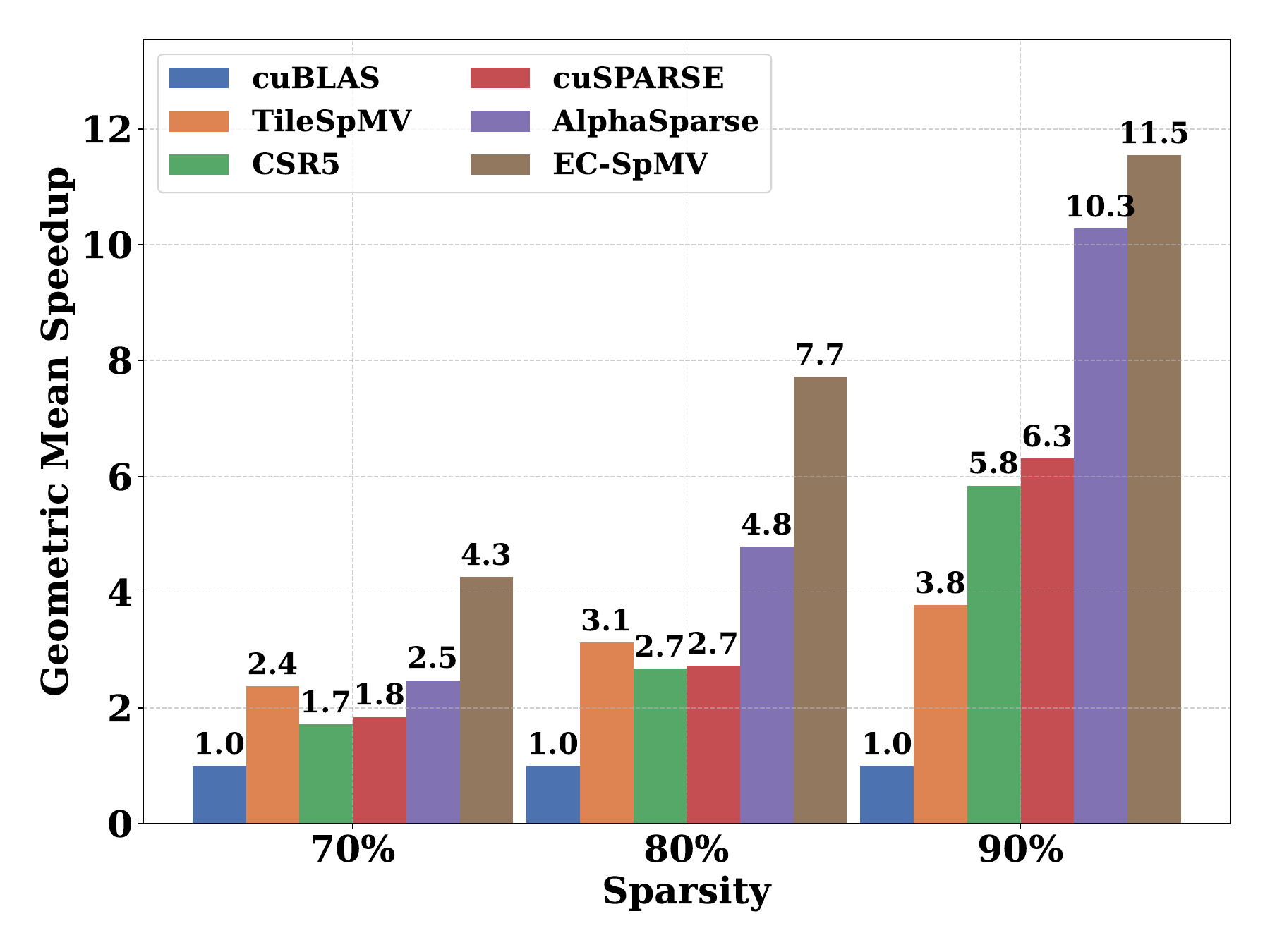}
        \caption{4090 for FP32 precision}
    \end{subfigure}
    
    \begin{subfigure}[]{0.33\textwidth}
        \centering
        \includegraphics[width=\textwidth]{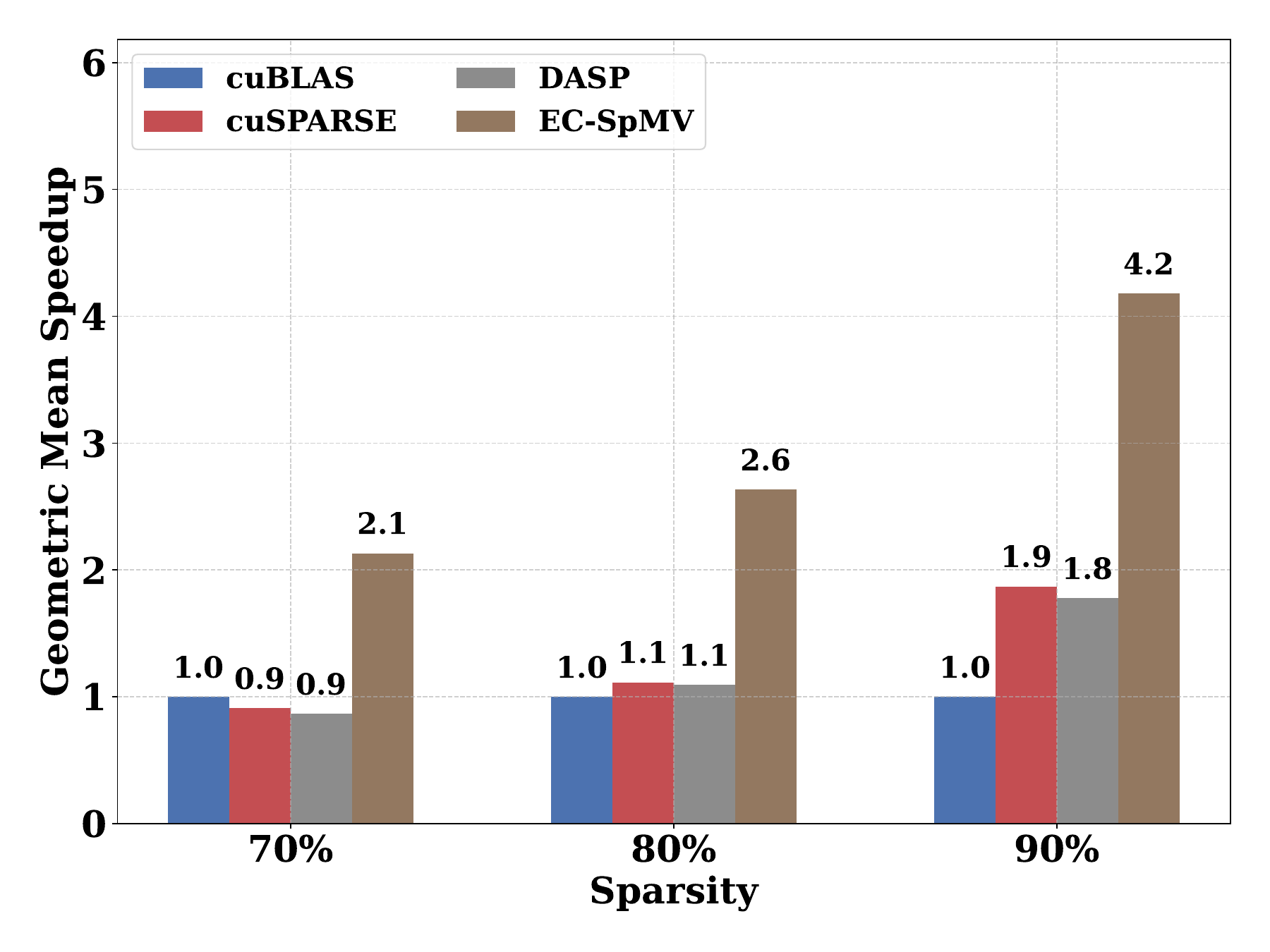}
        \caption{3080 Ti for FP16 precision}
    \end{subfigure}
    \begin{subfigure}[]{0.33\textwidth}
        \centering
        \includegraphics[width=\textwidth]{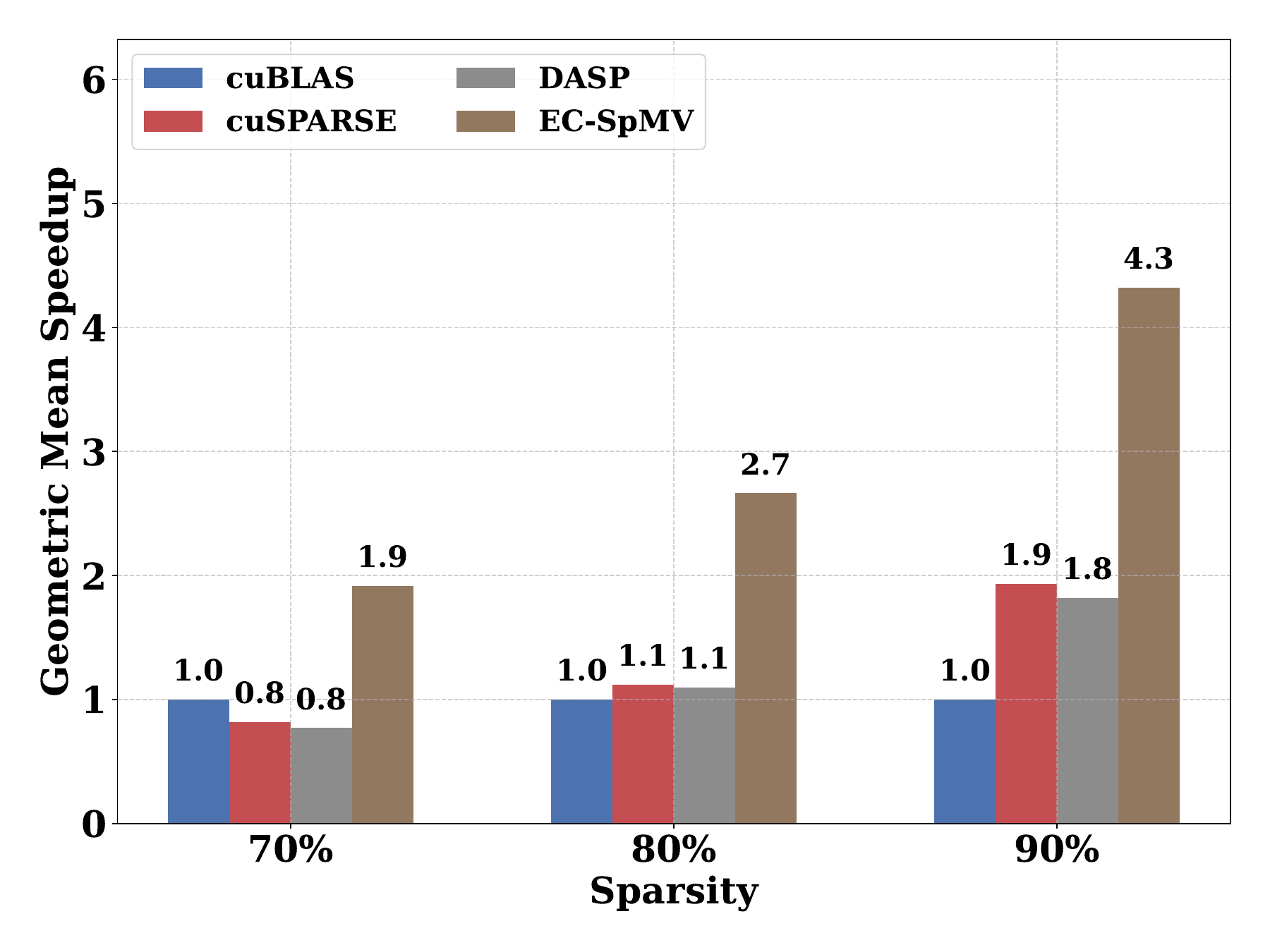}
        \caption{3090 for FP16 precision}
    \end{subfigure}
    \begin{subfigure}[]{0.33\textwidth}
        \centering
        \includegraphics[width=\textwidth]{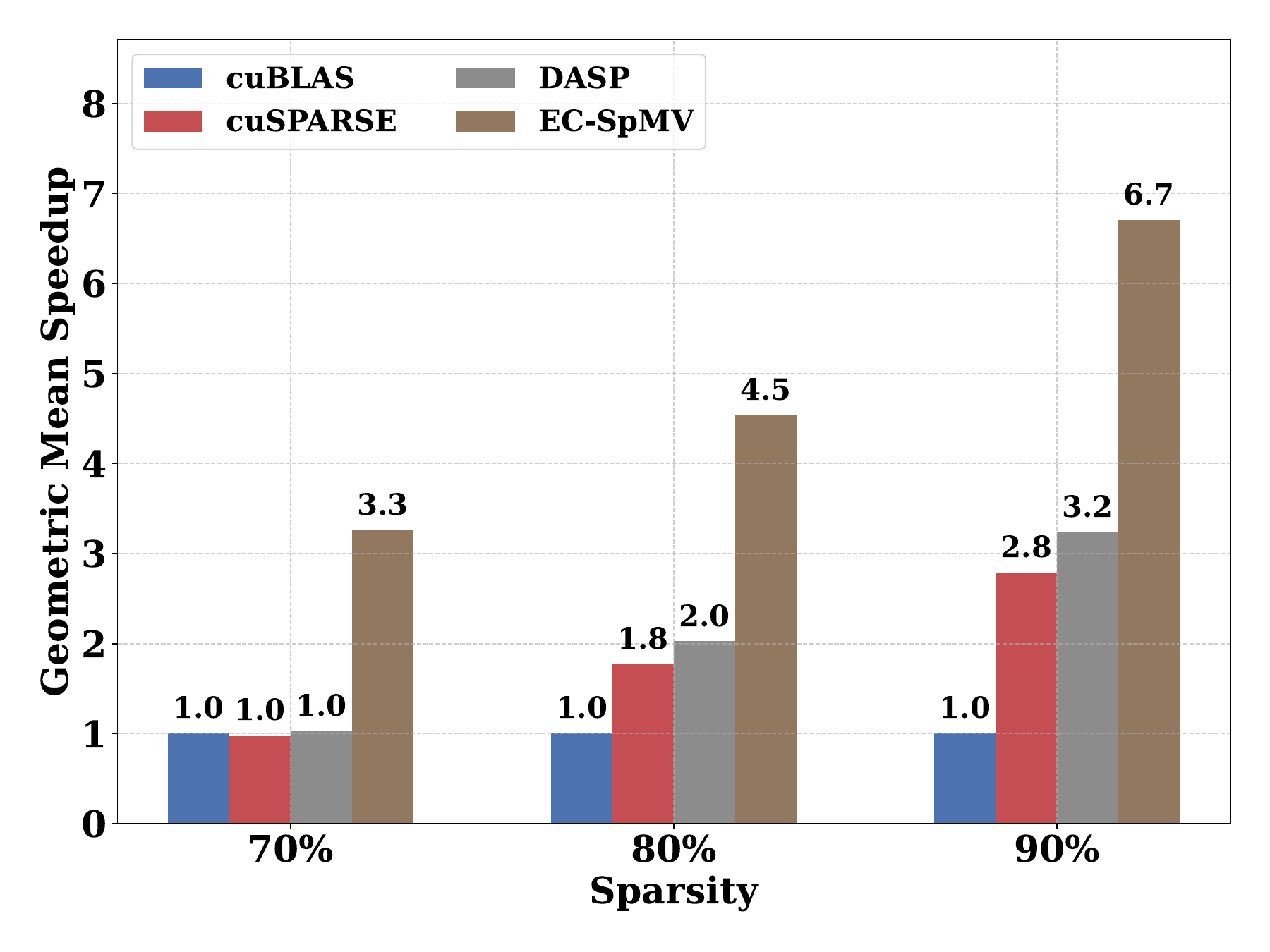}
        \caption{4090 for FP16 precision}
    \end{subfigure}
    \caption{The relative performance of various SpMV kernel implementations compared to cuBLAS.}
    \label{fig:kerperf}
\end{figure*}

In order to optimize data access, we employ double buffering to prefetch delta indices and block values.
While one set of data is being processed, the next set is fetched asynchronously, ensuring continuous data availability for computation (lines 28-29, lines 36-39).
The main loop iterates over \texttt{vector\_size} delta indices in the assigned block (lines 42-50). 
Within each iteration, each thread computes \(G\) partial results through the following steps:
(1) computing the current index by adding the delta index to the previous index (line 43),
(2) loading the corresponding input vector element (line 44), and
(3) multiplying this element by the \( g \) non-zero elements and accumulating the results in the \( g \) registers (line 47-49).

After processing all elements, a warp-level reduction is performed to aggregate the partial results from different threads within the warp (lines 54-56). 
Finally, the aggregated results are written to the output vector using atomic operations to safely handle concurrent updates across multiple warps and blocks (lines 59–63).

\section{Evaluation}
In this section, we present a comprehensive evaluation of {\toolName} to demonstrate its effectiveness across multiple dimensions.
We begin by analyzing the performance of the SpMV kernel and the effectiveness of the {\formatName} format.  
Subsequently, we conduct an ablation study to assess the effects of three optimizations: index compression, block extraction, and load balancing. Finally, we explore a real-world scenario by integrating {\toolName} into \textit{llama.cpp} \cite{llama.cpp} and measuring its end-to-end performance in LLM inference. All experiments were conducted on modern GPU hardware, with detailed specifications provided in Table \ref{tab:hardware}.

\begin{table}[h]
\caption{Hardware Configurations}
\label{tab:hardware}

\begin{tabular}{cccc}
   \toprule
   CPU (Intel Xeon)               & GPU                 & Memory\\ \midrule
Silver 4316 (20 cores, 2.3 GHz) & RTX 3080Ti (12 GB) & 256 GB  \\
Silver 4316 (20 cores, 2.3 GHz) & RTX 3090\ \ (24 GB)     & 256 GB   \\
W5-3435  (16 cores, 3.1 GHz)     & RTX 4090\ \ (24 GB)    & 768 GB    \\ \bottomrule
\end{tabular}%

\end{table}

To evaluate {\toolName}, we constructed a dataset of sparse matrices derived from various LLMs. Specifically, we leveraged SparseGPT \cite{frantar2023sparsegpt} to prune LLaMA2 \cite{touvron2023llama} and OPT \cite{zhang2022opt} models of various sizes (7B, 13B, 30B, and 70B) at sparsity levels of 70\%, 80\%, and 90\%. The resulting matrices, with dimensions ranging from 1024 $\times$ 8192 to 8192 $\times$ 28672, capture diverse sparsity patterns and structural characteristics, ensuring that our evaluation reflects realistic deployment scenarios.

\subsection{Kernel Performance}
In this section, we compare {\toolName} with several state-of-the-art solutions: cuSPARSE \cite{cusparse}, cuBLAS (Tensor Core enabled) \cite{cublas}, CSR5 \cite{liu2015csr5}, TileSpMV \cite{niu2021tilespmv}, AlphaSparse \cite{du2022alphasparse}, and DASP \cite{lu2023dasp}. 
\begin{itemize}
\item \textbf{cuSPARSE} and \textbf{cuBLAS} are NVIDIA's official libraries for sparse and dense operators, respectively.
\item \textbf{CSR5-based SpMV} employs a redesigned segmented sum algorithm to achieve high SIMD utilization.
\item \textbf{TileSpMV} divides sparse matrices into tiles, selecting the optimal sparse format for each tile.
\item \textbf{AlphaSparse} searches for the best sparse matrix storage format and SpMV implementation for a sparse matrix. Due to its high search overhead, we limit the search time to 10,000 seconds.
\item \textbf{DASP} utilizes Tensor Cores on GPUs to accelerate SpMV kernel performance.
\end{itemize}
Since some solutions support only specific precisions for storing non-zero values, we evaluate {\toolName} in FP32 against CSR5, TileSpMV, and AlphaSparse, and in FP16 against DASP. Performance comparisons with cuBLAS and cuSPARSE are presented for both precisions. The evaluations were conducted on three types of GPUs: NVIDIA GeForce RTX 3080 Ti, 3090, and 4090. All experiments were run with CUDA Toolkit version 12.2, except for TileSpMV, which was tested with version 11.1 due to compatibility issues.

The usage of hierarchical block extraction and the {\formatName} format in {\toolName} significantly enhances data locality and memory access efficiency, thereby substantially improving SpMV performance.
As shown in Figure \ref{fig:kerperf}, {\toolName} demonstrates impressive speedups across all GPUs and precisions. On the 3080 Ti, {\toolName} achieves geometric mean speedups ranging from $1.33\times$ to $4.45\times$ in FP32 and from $2.31\times$ to $2.86\times$ in FP16, compared to the baselines. 
On the 3090, the speedups range from $1.36\times$ to $4.52\times$ in FP32 and from $2.43\times$ to $2.80\times$ in FP16. 
On the 4090, speedups range from $1.46\times$ to $7.24\times$ in FP32 and from $2.45\times$ to $4.63\times$ in FP16. 
Notably, {\toolName} consistently outperforms cuBLAS in all cases, demonstrating its efficacy in accelerating LLM inference.

\begin{figure}[h]
    \centering \includegraphics[width=0.45\textwidth]{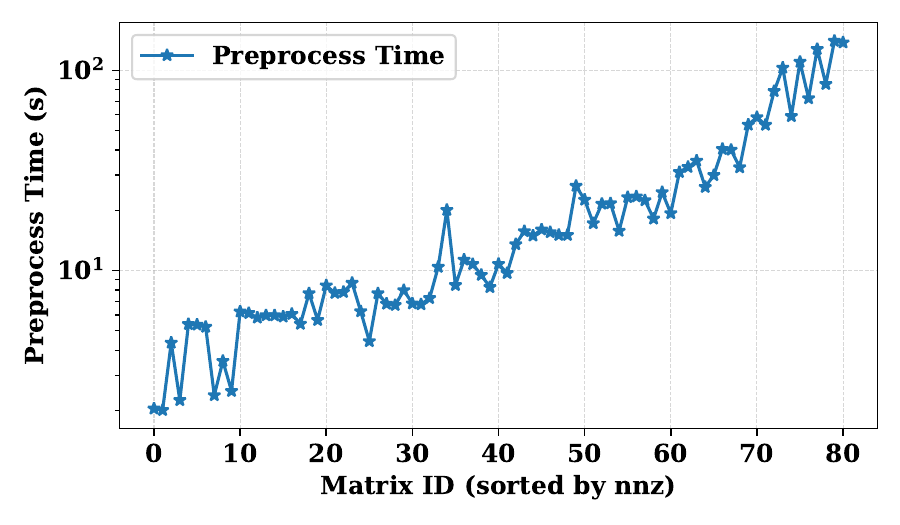}
    \caption{The preprocess overhead of {\toolName}.}
    \label{fig:p_cost}
\end{figure}
The preprocessing procedure of {\toolName} comprises two main stages: hierarchical block extraction and sparse matrix format conversion. As illustrated in Figure~\ref{fig:p_cost}, the preprocessing overhead remains below 100 seconds for most matrices. Notably, this cost is incurred only once during model deployment and can be amortized across multiple executions.

\subsection{Effectiveness of {\formatName}}
\begin{figure}[h]
    \centering
    \begin{subfigure}[]{0.23\textwidth}
        \centering
        \includegraphics[width=\textwidth]{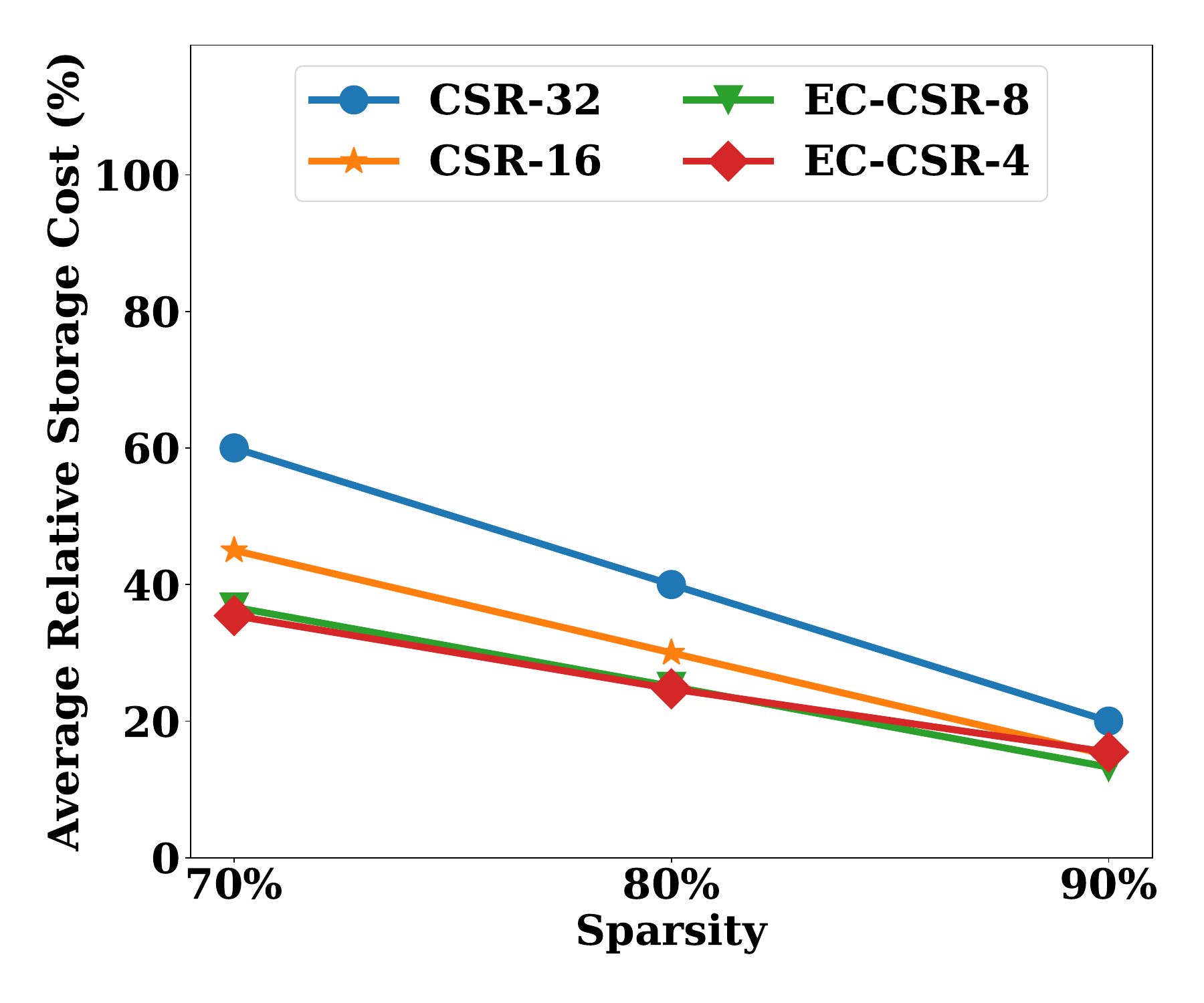}
        \caption{Storage in FP32 precision}
    \end{subfigure}
    \hfill
    \begin{subfigure}[]{0.23\textwidth}
        \centering
        \includegraphics[width=\textwidth]{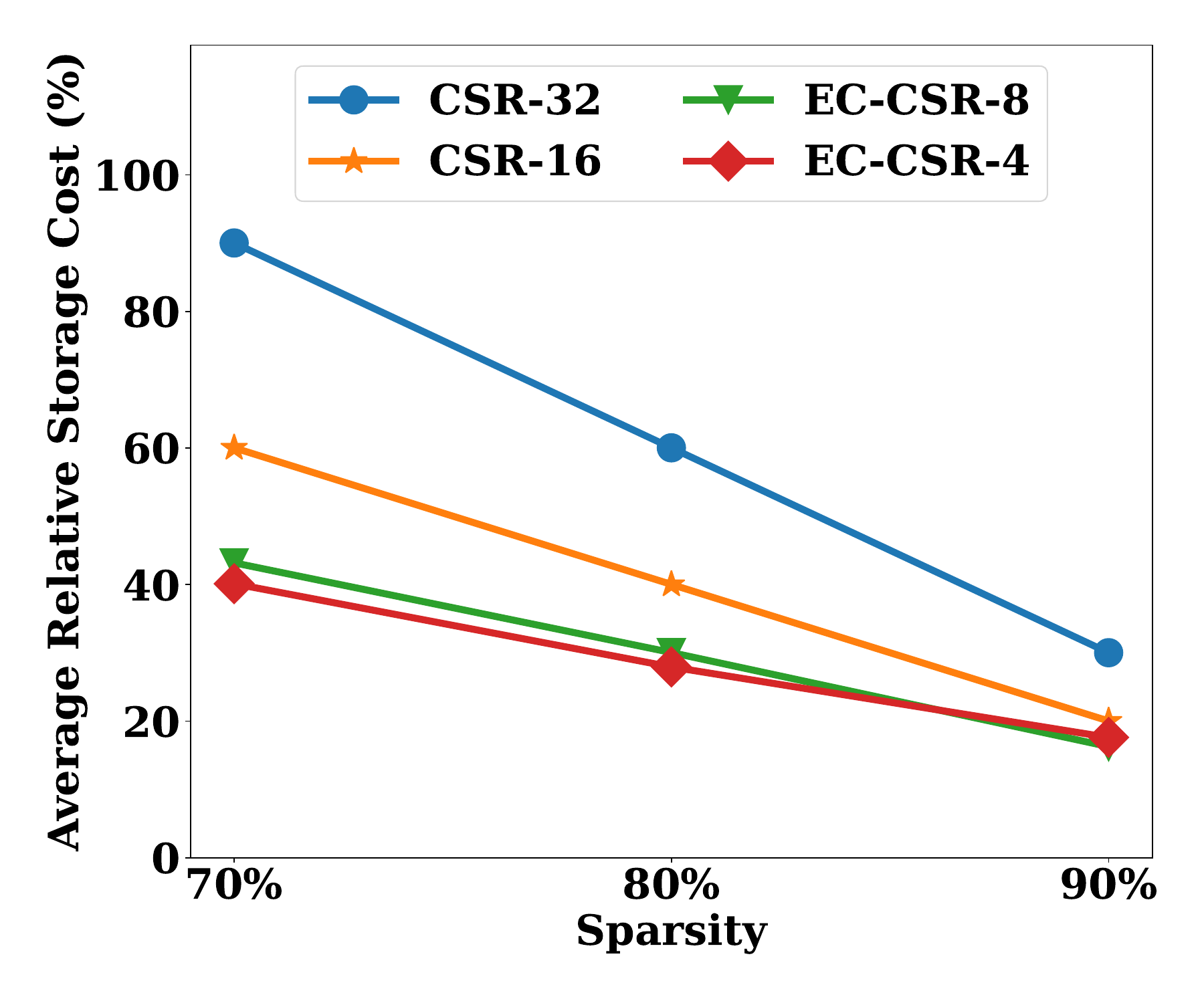}
        \caption{Storage in FP16 precision.}
    \end{subfigure}
    \caption{The relative storage overhead on different formats.}
    \label{fig:cost}
\end{figure}

Figure \ref{fig:cost} compares the storage overhead of different formats, using FP32 and FP16 precisions for non-zero values, relative to the dense format.
The ``X'' in CSR-X and {\formatName}-X represents the precision bit count for absolute and delta indices, respectively.
{\toolName} reduces storage overhead through two key mechanisms:
hierarchical block extraction, which reduces the number of indices; and index compression, which lowers the cost of each index by replacing absolute indices with delta indices. 
For FP32 precision, EC-CSR-8 achieves storage reductions of 38.9\%, 37.3\%, and 33.7\% compared to CSR-32 at three sparsity levels. Similarly, EC-CSR-4 achieves storage reductions of  40.9\%, 38.2\%, and 22.7\%. 
For FP16 precision, EC-CSR-8 achieves storage reductions of 52.0\%, 49.9\%, and 45.8\%, while EC-CSR-4 achieves storage reductions of 55.4\%, 53.5\%, and 41.3\%.

\begin{table}[b]
\centering
\begin{tabular}{cccc}
\toprule
DataTypes                  &70\%  & 80\% & 90\%  \\ \midrule
{\formatName}-4         & 6.35\% & 7.39\% & 30.43\%        \\ 
{\formatName}-8         & 0.75\% & 0.71\% & 1.87\%         \\ 
\bottomrule
\end{tabular}
\caption{The padding overhead at different precisions of delta indices for three sparsity levels.}
\label{tab:padding_cost}
\end{table}

Compared to EC-CSR-8, the EC-CSR-4 format, despite using half the index precision, fails to achieve significant storage reductions and even slightly increases storage requirements in some cases.
This behavior arises from the substantial padding overhead incurred to ensure that delta indices remain within the limited range of \texttt{uint4} precision, particularly at high sparsity levels. For instance, at 90\% sparsity, the padding overhead reaches 30.43\%, as detailed in Table \ref{tab:padding_cost}.
These findings suggest that co-designing pruning algorithms to explicitly control the distance between adjacent non-zero elements can further optimize delta indexing.

\subsection{Ablation Study}

\begin{figure}[h]
    \centering
    \includegraphics[width=0.5\textwidth]{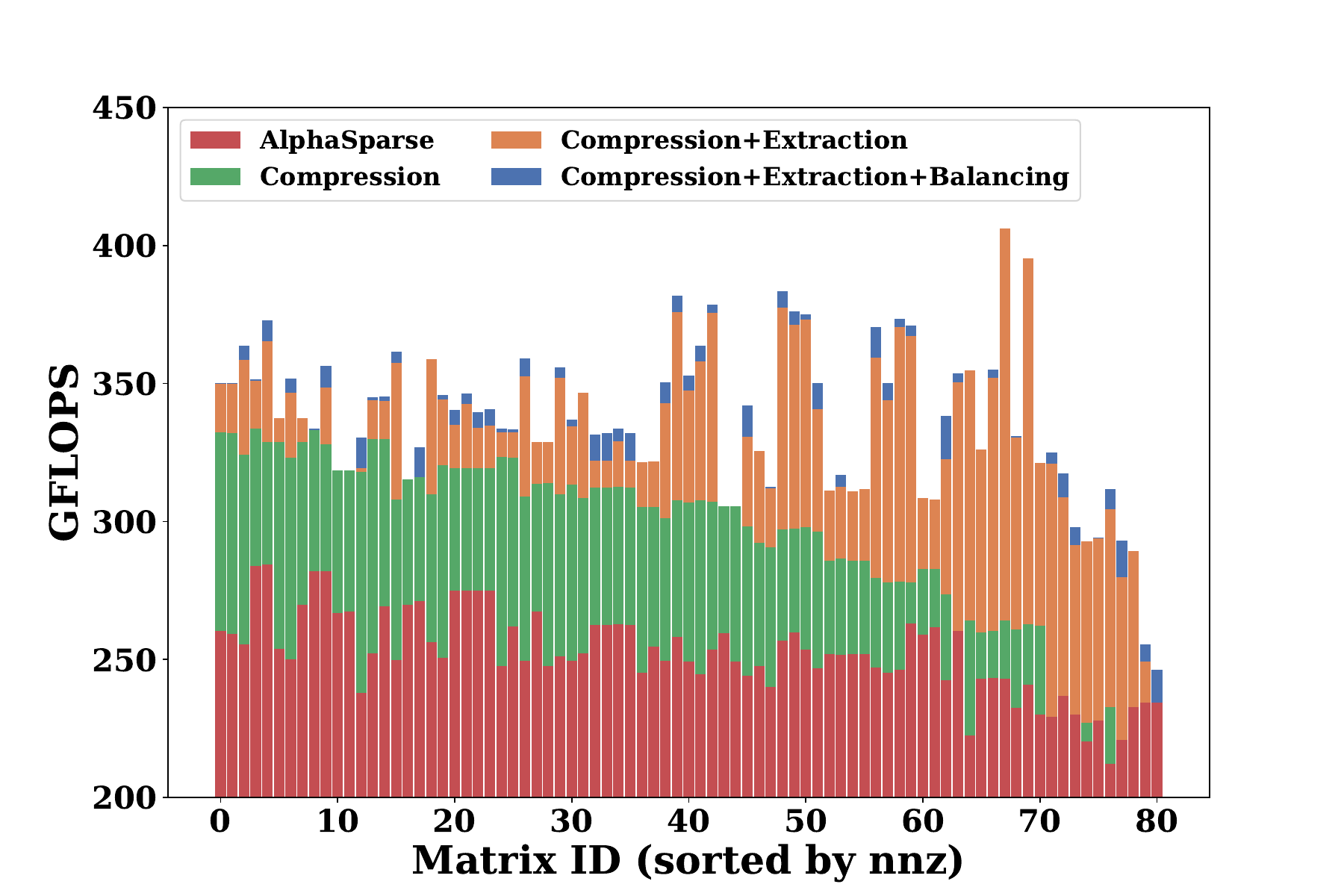}
    \caption{Effect of various optimizations.}
    \label{fig:ablation}
\end{figure}

Figure \ref{fig:ablation}  illustrates the contributions of various optimizations implemented in {\toolName} to SpMV kernel performance, taking FP32 precision on the NVIDIA 3080 Ti as an example. 
Among these, {\toolName} with only index compression improves memory access efficiency by replacing absolute indices with delta indices, yielding a 15.4\% performance gain over AlphaSparse, which automatically designs storage formats and generates optimized SpMV kernels.
Further gains are achieved through hierarchical block extraction, which enhances data locality by exploiting blocks at multiple granularities, resulting in an additional 15.7\% improvement over index compression alone. Additionally, the load balancing strategy can boost performance on certain sparse matrices by up to 10.12\%.

\begin{table}[b]
\centering
\begin{tabular}{cccc}
\toprule
Methods                 & 64  & 128 & 256  \\ \midrule
llama.cpp (\textbf{13.48 GB})         & 45.36 & 43.64 & 41.03        \\ 
llama.cpp+cuSPARSE (\textbf{12.19 GB}) & 44.55 & 42.49 & 39.93         \\ 
{\toolName} (\textbf{6.28 GB})   & 52.23 & 49.97 & 45.45          \\ \bottomrule
\end{tabular}
\caption{The number of tokens generated per second of three implementations under different prediction lengths.}
\label{tab:e2e}
\end{table}

\subsection{Case Study}

We applied the SparseGPT pruning method \cite{frantar2023sparsegpt} to achieve a 70\% sparsity level in the Llama2-7b model, resulting in a perplexity increase from 5.12 to 24.00 on the Wikitext dataset. 
The pruned model was deployed on an NVIDIA 3090 GPU using the \textit{llama.cpp} framework \cite{llama.cpp}. Our end-to-end experiments, shown in Table \ref{tab:e2e}, demonstrate that {\toolName} outperforms both the standard llama.cpp and its cuSPARSE integration in terms of inference metrics.
In terms of storage, the dense Llama2-7b-0.7 model uses 13.48 GB, while CSR-32 reduces this to 12.19 GB.  
{\formatName} further reduces storage to only 6.28 GB, enabling inference on GPUs with limited memory such as the RTX 3080 Ti. 
On the 3090 GPU, {\toolName} boosts performance by 15.1\%, 14.5\%, and 10.8\% across three prediction lengths, respectively. 
These results demonstrate the real-world applicability of {\toolName} for sparse LLM inference.
While this work focuses on kernel-level SpMV performance and storage efficiency, future research will explore broader end-to-end optimizations for sparse LLM inference.

\section{Related Work}
This section introduces related work on SpMV optimization, including tiling and sparse matrix storage formats.

Tiling \cite{aktulga2014optimizing,niu2021tilespmv,  hong_adaptive_2019, jiang2020novel,du2022alphasparse,wilkinson2023register, lin2024lo} is a method that divides a sparse matrix into tiles to enhance hardware utilization and data locality.
Aggressive tiling methods \cite{hong_adaptive_2019, jiang2020novel,zhao2025acc} further improve data locality by clustering and adjacent placing rows with a similar distribution of non-zero elements in sparse matrices. 
Our {\toolName} extends the aggressive tiling methods to hierarchical block extraction to extract as many blocks as possible.

Sparse matrix storage formats are critical for improving SpMV performance. Related studies include CSR-Adaptive \cite{greathouse2014efficient}, ACSR \cite{ashari2014fast}, CSR5 \cite{liu2015csr5}, ELL-R \cite{vazquez2011new}, CSB \cite{aktulga2014optimizing}, HYB \cite{bell2008efficient}, 
OSKI \cite{ vuduc2005oski}, TileSpMV \cite{niu2021tilespmv}, and AlphaSparse \cite{du2022alphasparse}.
These methods typically cater to high-sparsity matrices in scientific computing and are not optimized for sparse LLMs with relatively moderate sparsity. 
We design the {\formatName} format, which employs index compression and memory access optimization to reduce storage overhead and improve the performance of the SpMV kernel.
Moreover, the {\formatName} format is designed to handle the specific block structures extracted by the hierarchical block extraction method. 
Index compression refers to techniques for reducing the storage overhead \cite{freire2024leveraging}, including direct index compression, bitmap encoding \cite{fan2025spinfer}, delta-to-diagonal encoding \cite{bollhofer2002relations,sun2011optimizing}, and delta encoding \cite{kourtis2008optimizing,willcock2006accelerating, maggioni2013adell}.
Due to the irregular and moderate sparsity of the sparse weight matrices, the proposed {\formatName} uses delta encoding to compress the indices.

\section{Conclusion}
In this paper, we introduce {\toolName}, a novel method for accelerating SpMV kernels in sparse LLMs. {\toolName} leverages a hierarchical block extraction strategy that significantly enhances data locality, enabling more efficient memory access patterns tailored to LLM sparsity. 
Additionally, we introduce {\formatName}, a sparse matrix storage format designed for sparse LLMs, which cuts storage needs while maintaining memory efficiency. 
{\toolName} boosts performance by $2.43 \times$ in FP16 and $1.38 \times$ in FP32 precision over state-of-the-art SpMV kernels, and {\formatName} reduces storage overhead by 55.4\% compared to the CSR format. 
Future work will try to extend these techniques to other sparse neural network operators, such as Sparse Matrix-Matrix Multiplication (SpMM) and Sampled Dense-Dense Matrix Multiplication (SDDMM).

\bibliography{ref}

\end{document}